\documentclass{llncs}
\usepackage{listings}
\usepackage{hyperref}
\usepackage{graphicx}
\usepackage{amsmath}
\usepackage{color}
\usepackage{lineno}
\usepackage{multirow}

\title{Edit and Verify}
\author{Radu Grigore\inst{1} and Micha{\l} Moskal\inst{2}}

\institute{
UCD CASL, University College Dublin, Belfield, Dublin 4, Ireland
\and
Institute of Computer Science, University of Wroc{\l}aw,
ul.~Joliot-Curie 15, 50-383 Wroc{\l}aw, Poland,
\email{mjm@ii.uni.wroc.pl}
}
\date{}

\def\lstinlinen{\lstinline[basicstyle=\normalsize\sffamily]}

\newtheorem{DEF}{Definition}
\DeclareMathOperator{\preV}{in}
\DeclareMathOperator{\postV}{out}
\DeclareMathOperator{\Prune}{P}

\def\mobius{{\sc Mobius}}
\def\unsat{{\sc Unsat}}
\def\escjava{\hskip 0pt\hbox{ESC/Java2}}
\def\specsharp{\hskip 0pt\hbox{Spec$^\#$}}
\def\true{\top}
\def\false{\bot}
\def\equ{\Leftrightarrow}

\def\pmi{\Leftarrow}

\def\seq#1{#1_1, #1_2, \ldots}
\def\vec#1{{\bf #1}}

\begin{document}

\maketitle

\begin{abstract}
Automated theorem provers are used in extended static checking,
where they are the performance bottleneck. Extended static
checkers are run typically after incremental changes to the
code. We propose to exploit this usage pattern to improve
performance. We present two approaches of how to do so and 
a full solution.
\end{abstract}

%
%
%
%
%

\section{Introduction}

\begin{figure}[bh]
  \centering
  \includegraphics{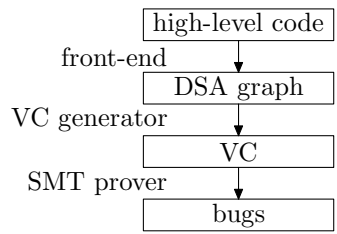}
  \caption{The architecture of an ESC}
  \label{fig:esc_arch}
\end{figure}

Extended static checking~\cite{escjava} is a technology
that makes automated theorem proving relevant to a wide group
of programmers. The architecture of an Extended Static Checker~(ESC)
is similar to that of a compiler (see Fig.~\ref{fig:esc_arch}).
It has a front-end that translates high-level code and specifications
into a simpler intermediate representation, and a back-end that 
formulates first order logic formulas as queries for a theorem prover.
The queries are called \emph{verification conditions}~(VCs).
If the ESC is sound then the VC is \unsat\ only if
the code meets its specifications; if the ESC is complete then
the program meets its specification only if the VC is \unsat.
\escjava~\cite{escjava} is an ESC that was designed to be unsound
and incomplete (as a tradeoff to make it more usable in practice);
\specsharp~\cite{boogie} is an ESC that was designed
to be sound. 

In this article we shall assume an ideal ESC that 
is both sound and complete. Automated first order theorem provers
used in extended static checking are incomplete: They either find a 
proof that a formula is \unsat\ or they give an assignment that 
\emph{probably} satisfies the formula. As a result, even if the 
ESC is sound and complete, spurious warnings are possible.

The purpose of an ESC is to provide warnings that help programmers
to write high-quality code. In practice it is used much like a
compiler. Either the programmer runs it periodically or the
Integrated Development Environment (IDE) runs it in the 
background. Because of these usage patterns,
performance is quite important. The bottleneck is the prover. Luckily,
the fact that the ESC is run often can be exploited since it means that
the program does not change much between two runs. Compilers already
exploit this by doing incremental compilation~\cite{schwartz1984icm}. 
ESCs do checking in a modular way, method by method. Nevertheless, 
once the contract of a method is altered all its clients must be 
rechecked. In such a scenario the VCs of the clients do not 
change much.

\begin{figure}[th]
  \centering
\begin{tabular}{c} 
\begin{lstlisting}
// blank line                           // (1)
class Day {
  //@ ensures 1 <= \result && \result <= 12;
  public abstract int getMonth();

  //@ ensures 1970 <= \result;
  //@ ensures \result <= 2038; 			// (2)
  public abstract int getYear();

  //@ ensures 1 <= \result && \result <= 31;
  public abstract int getDay();

  //@ ensures 1 <= \result;
  //@ ensures \result <= 366; 			// (3)
  public int dayOfYear() {
    int offset = 0;
    if (getMonth() > 1) offset += 31;
    if (getMonth() > 2) offset += 28;
    if (getMonth() > 3) offset += 31;
    if (getMonth() > 4) offset += 30;
    if (getMonth() > 5) offset += 31;
    if (getMonth() > 6) offset += 30;
    if (getMonth() > 7) offset += 31;
    if (getMonth() > 8) offset += 31;
    if (getMonth() > 9) offset += 30;
    if (getMonth() > 10) offset += 31;
    if (getMonth() > 11) offset += 30;
    boolean isLeap = getYear() % 4 == 0 && 
    		     (getYear() % 100 != 0 || getYear() % 400 == 0);
    //@ assert offset <= 335; 			// (4)
    if (isLeap && getMonth() > 2) offset++;
    return offset + getDay();
  }
}
\end{lstlisting}
\end{tabular}
  \caption{Typical evolution of annotated Java code}
  \label{fig:java_evol}
\end{figure}

This paper (1)~argues for the importance of using techniques 
analogous to incremental compilation in software verification,
(2)~formalizes the problem and explores possible solutions 
(Sect.~\ref{sec:discussion}), (3)~presents a specific solution
that works exclusively inside an automated theorem prover
(Sect.~\ref{sec:prune}), in the process (4)~presents a
technique to heuristically determine similarities between
formulas, and~(5) gives a mechanically verified proof
for the correctness of a part of the specific solution 
presented.

\section{Discussion and Definitions}
\label{sec:discussion}

The problem in a nutshell is how to do incremental extended static
checking. We shall explore the solution space and then we will
see in detail a particular solution, including some experimental
data.

Consider the JML-annotated Java code from Fig.~\ref{fig:java_evol} 
When checking the method 
\lstinlinen|dayOfYear| the ESC will assume 
the implicit empty precondition holds and will try to prove the postcondition.
It will also try to prove all the explicit and implicit assertions 
in the body. When the method \lstinlinen|getMonth| is called the
ESC inserts (implicit) assertions for its preconditions followed
by assumptions for its postconditions. Moreover, the ESC will
introduce assertions that ensure the absence of runtime exceptions.
For example, the receiver object of a method call is asserted
to be nonnull.

Notice the lines marked by (1), (2), (3), and (4). Adding these
lines represents typical edits that can be done on annotated source code.
For example, line \lstinlinen|(3)| is a newly added postcondition.
An incremental VC would only check if this new assertion
holds, provided that the last VC was \unsat. It is somehow cumbersome to
formulate the problem precisely at the source code level. We can be
more precise by descending at the level of an idealized
intermediate representation, a \emph{Dynamic Single Assignment~(DSA)
graph}.

\begin{DEF}[DSA graph]
The \emph{DSA graph} of a method is a directed acyclic (control 
flow) graph. Its vertices are $1, 2, \ldots$ and they
are labeled respectively by the first order logic formulas
$\seq\phi$. A vertex represents either an \emph{assertion}
(in which case we say it is \emph{black}) or an \emph{assumption}
(in which case we say it is \emph{white}). 
We denote the set
of vertices that are predecessors of $v$ by $\preV(v)$ and the
set of successors of $v$ by $\postV(v)$. The \emph{in-degree}
of $v$ is $|\preV(v)|$ and the \emph{out-degree} is $|\postV(v)|$.
The nodes with in-degree zero are called \emph{initial nodes};
the nodes with out-degree zero are called \emph{final nodes}.
\end{DEF}

\noindent
The assertions model the postconditions of the verified method 
and the checks inside its body (such as the check that an index 
in an array access is in-bounds, a receiver of a method call is 
nonnull, the preconditions of a called method hold, explicit
JML assertions, and so on). The assumptions model postconditions 
of the called methods and semantics of the Java language 
(including properties ensured by the type system). 

For this presentation we simply assume that the intermediate
representation is obtained from the source code by some technique,
without committing to any one in particular. The curious reader 
can start exploring the subject from other 
papers~\cite{boogie,barnett2004voo,leino2005wpu,darvas2005rmc}.

The VC is generated from the intermediate
representation. The particular algorithm used has a big impact
on performance~\cite{flanagan2001aee,leino2005wpu}. Here we
only present a conceptually simple technique that illustrates 
well the general form VCs have in practice.

\begin{DEF}[behaviors]
Vertices have associated \emph{preconditions} denoted by
$\seq\alpha$, \emph{postconditions} denoted by $\seq\beta$, and
\emph{wrong behaviors} denoted by $\seq\gamma$ For all $i$ we have

\penalty+100

\begin{align}
\alpha_i &= 
  \begin{cases}
    \true &\text{for initial nodes} \\
    \bigvee_{v_j\in\preV(v_i)} \beta_j &\text{for non-initial nodes}
  \end{cases} \\
\beta_i &= \alpha_i\land\phi_i \\
\gamma_i &= 
  \begin{cases}
    \alpha_i\land\lnot\phi_i &\text{for assertions} \\
    \false &\text{for assumptions}
  \end{cases}
\end{align}
\end{DEF}

\begin{DEF}[verification condition]
The \emph{verification condition} is
\begin{equation}
\psi=\bigvee_i \gamma_i
\end{equation}
\end{DEF}

\noindent
The wrong behaviors are something we want to avoid, therefore
we ask the prover if all the wrong behaviors are impossible
which is the same as asking if the VC is \unsat.  If it is, then the
ESC concludes that all the assertions are valid and the method is
correct. The basic idea behind the more efficient techniques
of generating VCs is to generate factored form.

\begin{table}[th]
  \centering
  \begin{tabular}{|c|c|c|}
    \hline
    Old & New & Simplified \\
    \hline
    \hline
    \includegraphics{main.1} & 
      \includegraphics{main.2} & 
      \includegraphics{main.3} \\
    \hline 
    $\psi_1=\phi_1\land\lnot\phi_2$ &
      $\psi_2=\left(\phi_1\land\lnot\phi_2\right)\lor
        \left(\phi_1\land\phi_2\land\lnot\phi_3\right)$ &
      $\psi'_2=\phi_1\land\phi_2\land\lnot\phi_3$ \\
    \hline
  \end{tabular}
  \caption{Simplification example}
  \label{tbl:simpl_ex}
\end{table}

The problem can now be stated as follows: Given two similar
formulas $\psi_1$ and $\psi_2$, find a formula $\psi'_2$ that
is \unsat\ if and only if $\psi_2$ is \unsat, provided that $\psi_1$
is \unsat. An example is given in Table~\ref{tbl:simpl_ex}.
The following equations show step by step how to compute~$\psi_2$
from its corresponding DSA graph.

\begin{align}
\alpha_1&=\true& \beta_1&=\phi_1&  \gamma_1&=\false \\
\alpha_2&=\phi_1& \beta_2&=\phi_1\land\phi_2& 
  \gamma_2&=\phi_1\land\lnot\phi_2 \\
\alpha_3&=\phi_1\land\phi_2& \beta_3&=\phi_1\land\phi_2\land\phi_3&
  \gamma_3&=\phi_1\land\phi_2\land\lnot\phi_3
\end{align}

\noindent
To make the example concrete the reader might wish to plug in
$\phi_1=x>2$ and $\phi_2=x>1$ and $\phi_3=x>0$.

Note that $\psi'_2=\phi_1\land\lnot\phi_3$ is sound too, but we
do not want to drop parts of the formula that are assumptions
because they can make the proof easier. The simplified formula 
can be obtained in
two ways. One is to replace the assertions that appear in both
DSA graphs by assumptions and generate the VC for the modified
DSA graph; the other is to work directly on the formulas 
$\psi_1$ and $\psi_2$. In this paper we will explore in greater 
detail the latter. 

In both approaches, a solution has to solve
two subproblems. First, we must find a correspondence between
parts of the two DSA graphs (or formulas). Second, we must
simplify one of the DSA graphs (or formulas). The methods we
present in the next section for finding a correspondence
between parts of the formulas can be partially reused for
finding a correspondence between parts of the DSA graphs.
Simplifying a formula is harder than changing assertions into
assumptions, but on the other hand it is independent of the 
particular intermediate representation used.

\section{Pruning First Order Formulas}
\label{sec:prune}

One subproblem is to find a correspondence between parts
of $\psi_1$ and parts of~$\psi_2$. We substitute (some) 
uninterpreted constants in $\psi_1$ by uninterpreted 
constants that appear in~$\psi_2$. We also normalize
the formulas with respect to commutative operators
(Fig.~\ref{fig:alg_sort}). We also use 
hash-consing~\cite{hashcons_old,hashcons_ml} so later terms 
are simply compared by reference equality.

Note that if $\psi_1$ is \unsat, then any
substitution that renames uninterpreted constants leaves it \unsat.
The only assumption we make in solving the second subproblem
is that $\psi_1$ is \unsat, so there is no `right' or `wrong'
correspondence between old and new constants. It is true, 
however, that for different substitutions of constants we 
will end up with different results~$\psi'_2$, some bigger 
and some smaller. Also we need to remember not to rename 
interpreted constants (such as~$1$ and~$42$).

\begin{figure}[t]
  \centering
\begin{tabular}{c}
\begin{lstlisting}
class Term
  public Name : string
  public Children : list[Term]
def SortTerm(t)
  def CompareTerms(a, b)
    def nc = a.Name.CompareTo(b.Name)
    if (nc != 0) nc
    else LexicographicCompare(a.Children, b.Children, CompareTerms)
  def children = t.Children.Map(SortTerm)
  if (IsCommutative(t)) Term(t.Name, t.Children.Sort(CompareTerms))
  else                  Term(t.Name, children)
def oldVC = SortTerm(oldVC)
def newVC = SortTerm(newVC)
\end{lstlisting}
\end{tabular}
  \caption{Normalizing queries}
  \label{fig:alg_sort}
\end{figure}

Assuming that all constants that are `the same' have the 
same name in $\psi_1$ as in $\psi_2$ would not allow us 
to prune the VC (to~$\false$) when the programmer only 
renamed a variable. (Variables in the program appear as
uninterpreted constants in the VC.) Even worse, the ESC encodes extra 
information in identifiers~\cite{leino2005get} that changes, 
for example, when a new line is added to the source Java file. 
Despite these variations, a human that sees both $\psi_1$ and 
$\psi_2$ is generally able to say which sub-term corresponds 
to which sub-term. So there are good chances to find a heuristic 
that works well!

We only consider renaming of uninterpreted constants because 
of the particular algorithm used to build VCs. If some of the 
function symbols would also need to be renamed, the algorithm 
can be easily extended by the standard technique of introducing 
a special function symbol $apply$, and replacing 
$f(t_1,\dots,t_n)$ with $apply(f,t_1,\dots,t_n)$.

The heuristic we use to find a good substitution assigns 
a \emph{similarity} value to each pair of (old, new) constants 
and then finds a maximum bipartite matching (using the Hungarian
method~\cite{hungarian_alg}) between the old and
the new constants. A complete bipartite graph is 
constructed from the set~$V_1$ of uninterpreted constants that appear
in~$\psi_1$ and the set~$V_2$ of uninterpreted constants that appear
in~$\psi_2$. Each pair $(i, j)\in V_1\times V_2$ 
has an associated weight, which in this case is
the similarity of the two constants. A matching
is a subset $M\subset V_1\times V_2$ such that
for all pairs $(i, j)\in M$ and $(i', j')\in M$
we have $ i = i'$ if and only if $ j = j' $.
The weight of the matching is the sum of the weights of all its elements.
The similarity has two components: One is the length of the longest
common subsequence~\cite{lcs} of the two identifiers; the other, more
important, is how many times the constants appear in similar positions
in the two VCs.  

To measure similarity of position we use path
strings~\cite{termIndexing}.  A \emph{path string} is a sequence of
function symbols interleaved with the positions, on a path from the
root of the term to a particular occurrence of a sub-term.  For example
$f.2.g.1$ is a path string for the occurrence of~$b$ in $f(a,g(b,c))$, and
$f.2.g.2$ is a path string for~$c$.  We construct a \emph{stripped path string}
by treating logical connectives as function symbols, the entire formula as
a term, and skipping positions for commutative symbols.
For example $\wedge . \!\!\vee\!\! . f.2.g.1$ is the stripped path string 
for~$b$ in $(f(a,g(b)) \vee g(c)) \wedge g(d)$.
The \emph{environment} of a constant~$c$ in a formula $\psi$
is the multiset of the stripped path strings for all occurrences 
of~$c$ in~$\psi$.  Let~$E_1$ be the environment of~$x$ in~$\psi_1$ 
and~$E_2$ be the environment of~$y$ in~$\psi_2$. The similarity
of~$x$ and~$y$ is $2 |E_1 \sqcap E_2| - | (|E_1| - |E_2|) |$,
where $\sqcap$ is multiset intersection.
Other measures, that take environments into account, are also possible.

\begin{figure}[t]
  \centering
\begin{tabular}{c}
\begin{lstlisting}
def Prune(p1 : list[list[Term]], p2 : Term)
  def p1 = Flatten(p1)
  // |p1| is a DNF form, assumed to be UNSAT
  match (p2.Name)
    | "and" =>
        mutable common = []
        foreach (x in p1) foreach (y in x) common = y :: common
        def p1 = p1.Map(x => x.Filter(y => !common.Contains(y)))
        def p2 = p2.Children.Filter(y => !common.Contains(y))
        if (p1.Contains([])) Term("false", [])
        else                 Term("and", common + p2.Map(x => Prune(p1, x)))
    | "or" =>
        Term("or", p2.Children.Map(x => Prune(p1, x)))
    | _ =>
        if (p1.Exists(x => Implies(p2, Term("and", x)))) Term ("false", [])
        else                                             p2
def prunedVC = Prune([[oldVC]], newVC)
\end{lstlisting}
\end{tabular}
  \caption{Pruning the VC}
  \label{fig:alg_prune}
\end{figure}

The algorithms are presented as Nemerle-like pseudocode~\cite{nemerle}.
Some obvious optimizations are omitted\footnote{See
\url{http://nemerle.org/svn.fx7/branches/fx8/Pruner.n} for all details.} 
to improve readability. We also omit textbook algorithms.
The algorithm for normalizing queries with respect to 
commutative operators is given in Fig.~\ref{fig:alg_sort}.
It recursively sorts arguments of commutative operators
using lexicographic ordering.

The second subproblem, simplification of formulas, is
solved by the pruning algorithm in Fig.~\ref{fig:alg_prune}. 
The function \lstinlinen|Prune| returns a formula equisatisfiable
to \lstinlinen|p2| under the assumption that all elements
of \lstinlinen|p1| are \unsat. Elements of \lstinlinen|p1|
are conjunctions represented as lists.

The function \lstinlinen|Implies| explores the structure of two
formulas and returns \lstinlinen|true| only if the first is
stronger than the second. The last branch is clearly correct: 
If \lstinlinen|p2| is stronger than a conjunct known to be 
\unsat\ then it is also \unsat. In the case that 
\lstinlinen|p2| is a disjunction we can treat  its children 
independently. The case when \lstinlinen|p2| is
a conjunction is more interesting. To understand why it works
consider a small example.

\begin{align}
\psi_1 &= (\phi_1\land\phi_2) \lor (\phi_3\land\phi_4) \\
\psi_2 &= \phi_2 \land \phi_4 \land (\phi_1\lor\phi_3) \\
\psi'_2&= \phi_2 \land \phi_4 \land \false = \false
\end{align}

\noindent
We write $\Prune(\psi_1,\psi_2)=\psi'_2$ for the result of
pruning $\psi_2$ under the assumption that $\psi_1$ in \unsat.
The common part of $\psi_1$ and $\psi_2$, as computed in the
variable \lstinlinen|common| in Fig.~\ref{fig:alg_prune},
is $\phi_2\land\phi_4$. Pruning $\phi_1\lor\phi_3$ knowing 
that $\phi_1\lor\phi_3$ is \unsat\ results in~$\false$. 
The formulas that appear in both $\psi_1$ and $\psi_2$ 
can always be factored.

\begin{align}
     & (\phi_1\land\phi_2)\lor(\phi_3\land\phi_4) \\
\pmi & (\phi_1\land\phi_2\land\phi_4)\lor(\phi_3\land\phi_2\land\phi_4) \\
\equ & \phi_2\land\phi_4\land(\phi_1\lor\phi_3)
\end{align}

\noindent
Hence, we can always reduce the problem to the form

\begin{align} 
\psi_1 &= \phi'_1\land\phi'_2 \\
\psi_2 &= \phi'_1\land\phi'_3 \\
\psi'_2 &= \phi'_1\land\Prune(\phi'_2,\phi'_3)
\end{align}

\noindent
where $\phi'_1$ is the common part and $\phi'_2$ is what we
assume to be \unsat\ while pruning~$\phi'_3$ (see also Fig.~\ref{fig:alg_prune}). 
In this example $\phi'_1=\phi_2\land\phi_4$ and 
$\phi'_2=\phi'_3=\phi_1\lor\phi_3$.
It is easy to see that the above is correct, by doing a case
analysis on whether $\phi'_1(\vec x)$ holds for some vector~$\vec x$. 
The formalization\footnote{Available at
\url{http://radu.ucd.ie/hp/papers/ev.html}} in Coq~\cite{coq}
of a simplified version of the pruning function emphasizes
the main points of the proof. The formulas abstract theories
by arbitrary predicates over the domain of uninterpreted
constants.

\begin{tabular}{c}
\begin{lstlisting}
Inductive Formula : Type :=
  | FPred : (Dom -> Prop) -> Formula
  | FAnd : Formula -> Formula -> Formula
  | FOr: Formula -> Formula -> Formula.
Fixpoint Eval (f : Formula) (x : Dom) {struct f} : Prop :=
  match f with
    | FPred p => p x
    | FAnd fa fb => Eval fa x /\ Eval fb x
    | FOr fa fb => Eval fa x \/ Eval fb x
  end.
\end{lstlisting}
\end{tabular}

\noindent
The simplified version of the algorithm whose proof we check
mechanically is

\begin{tabular}{c}
\begin{lstlisting}
Fixpoint Prune (p1 p2 : Formula) {struct p2} : Formula :=
  match p1, p2 with
    | FAnd a b, FAnd aa c => if eq a aa then FAnd a (Prune b c) else p2
    | _, FOr a b => FOr (Prune p1 a) (Prune p1 b)
    | _, _ => if eq p1 p2 then FPred PFalse else p2
  end.
\end{lstlisting}
\end{tabular}

\noindent
This function has two important invariants.

\begin{tabular}{c}
\begin{lstlisting}
Lemma PruneInvA : forall p1 p2 : Formula, forall x : Dom,
  (~ Eval p1 x -> Eval p2 x -> Eval (Prune p1 p2) x).
Lemma PruneInvB : forall p1 p2 : Formula, forall x : Dom,
  (~ Eval p1 x -> Eval (Prune p1 p2) x -> Eval p2 x).
\end{lstlisting}
\end{tabular}

\noindent
These are proved by double induction on the structure of 
\lstinlinen|p1| and \lstinlinen|p2|. We use one extra 
fact.

\penalty+50

\begin{tabular}{c}
\begin{lstlisting}
Lemma UnsatImp : forall a b : Formula,
  (forall x : Dom, Eval a x -> Eval b x) -> Unsat b -> Unsat a.
\end{lstlisting}
\end{tabular}

\penalty-100\noindent
At this point we can prove that the algorithm 
is sound and complete.

\begin{tabular}{c}
\begin{lstlisting}
Lemma PruneSound : forall p1 p2 : Formula, 
  Unsat p1 -> Unsat (Prune p1 p2) -> Unsat p2.
Lemma PruneComplete : forall p1 p2 : Formula, 
  Unsat p1 -> Unsat p2 -> Unsat (Prune p1 p2).
Theorem PruneCorrect : forall p1 p2 : Formula, 
  Unsat p1 -> (Unsat p2 <-> Unsat (Prune p1 p2)).
\end{lstlisting}
\end{tabular}

The algorithm in Fig.~\ref{fig:alg_prune} is more efficient
since it exploits the associativity and commutativity of the 
$\land$ and $\lor$ operators. The worst case time complexity 
is $O(mn)$, and arises when the formula known to be \unsat\
and the formula to be simplified have, respectively, the form

\begin{align}
\psi_1 &= \bigvee(\phi_1,\ldots,\phi_{m-1}) \\
\psi_2 &= \underbrace{\land\ldots\land}_{n\text{ times}} \phi_m
\end{align}

\noindent
where $\land$ and $\lor$ are written as $n$ary operators.
Unfortunately, the average case that
appears in practice is hard to describe. Experimental data
from $20$ cases suggests that the running time grows linearly
with the size of the formulas. But we need more data before 
we can make a definite statement 
(see Sect.~\ref{sec:related} for details).

\section{Case Study}

In this section, we explain how the common way of editing programs
affects the DSA and therefore also the VC and how pruning exploits
the changes.

Let us again consider the program from Fig.~\ref{fig:java_evol}.
We used \escjava\ to generate VCs for a version without any of the 
lines marked (1), (2), (3), and (4). This was the base case. Next 
we ran it on a method with only line (1) added, only line (2) added 
and so forth. Finally
we ran the pruning algorithm with the old formula being the base
case and the new formula being being VC for a method with an added line.
Table~\ref{tbl:case_study_res} lists three times for each such
formula. The first is the time it takes to prove the formula
using Simplify~\cite{simplify}; the second is the time it takes
to prune the formula; the third is the time it takes to prove
the pruned formula. The reader can note that the running times of 
Simplify on the original formulas vary rather nondeterministically.
In particular, one would expect the base case and the one with
an added empty line to have the same running time, but they do not.
The reason for this is a ``butterfly effect'' in the prover, where for
example a slight change in the selection of a literal for a case split
can cause large changes in the final shape of the proof search tree.

\begin{table}
\centering
\[
\begin{array}{|c|l|r|r|r|r|r|}
\hline
\mbox{\bf Marker} &
\mbox{\bf Description} &
\mbox{\bf Original} &
\mbox{\bf Pruning} &
\mbox{\bf Pruned}
\\
\hline
& \mbox{base case} &     		    20.91s &   &   \\  
(1) & \mbox{empty line} 	 	    & 17.59s  & 2.23s & 0.01s   \\
(2) & \mbox{irrelevant postcondition} & 16.91s & 2.31s & 0.06s  \\
(3) & \mbox{additional postcondition} & 21.65s & 2.19s & 19.34s \\
(4) & \mbox{assertion in the middle}  & 22.81s & 2.16s & 7.67s  \\
\hline
\end{array}
\]
  \caption{Case study results}
  \label{tbl:case_study_res}
\end{table}

The first edit operation (marked by \lstinlinen|(1)|) is adding an empty
line somewhere, or in general changing the locations of symbols.
As ESCs often use location information for encoding symbol
names, the uninterpreted constants in the second VC are different
than in the first one. Our algorithm generates a query that is 
just~$\false$.

The second edit strengthens the postcondition
of a method \lstinlinen|getYear| used in the verified
\lstinlinen|dayOfYear| method. Here, we are able to prune 
almost everything, i.e. the resulting query is propositionally 
\unsat.

The third edit adds a postcondition to the verified method.
We can imagine that the DSA graph gets one more black node
at the end, so this is the only thing that should be verified
now. In this case we do prune parts of the formula, it however
fails to speed up checking.

Finally the last edit adds an assertion near the end of the method.
Here the heuristics work well and the time is reduced considerably.

The \lstinlinen|dayOfYear| method (Fig.~\ref{fig:java_evol}) is 
an example of a case where the VC is relatively small (around 
$60$~kilobytes), but hard to prove. This is due to the large 
number of possible paths in the method.
There are other reasons methods can be hard to prove: methods can be more complicated,
the specifications can be complicated, the modelling of the language can
be more accurate (for example in multi-threading programs). All those
scenarios are good for our pruning algorithm as it runs in polynomial
time and can potentially save a lot of proving time. The bad case is
when the formula is large, but not that hard to prove. In particular it
sometimes happen that most of the time is spent just reading/writing
the formula and doing basic preprocessing, like skolemization.

\section{Related and Future Work}
\label{sec:related}

The work presented here parallels the work done in the
compiler community under the name \emph{incremental compilation}.
In the context of software verification by theorem
proving the term \emph{incremental verification} is
taken---it refers to the process of proving stronger
assertions using weaker ones as lemmas~\cite{uribe2000cmc}. 
Hence, we use the distinct term \emph{edit and verify} 
for the related idea of proving only what has not been
proven before, and doing so automatically. In the context
of interactive theorem proving the term \emph{proof reuse}
is used for a similar technique~\cite{proof_reuse}.

A Program Verification Environment (PVE) is the same for an ESC,
as an Integrated Development Environment (IDE) is for a compiler.
It provides an easy to use interface to the tool. As incremental
compilation is very useful in IDEs, we expect Edit and Verify
to be even more useful in PVEs. This is because static verification
consumes much more resources than compilation. There is
much research on software verification using PVEs, there is also
vast amount of interest from the industry in PVEs.

One of the goals of the \mobius\ research project~\cite{mobius}
is to produce a PVE for Java. Penelope~\cite{guaspari1990fva} is an early PVE that
processes a subset of Ada. Its designers chose to rely
on interactive theorem proving. The KeY Tool~\cite{KeyBook2007}
is a modern PVE for Java that uses the same approach but differs
in the mechanisms and theory of verification condition generation.
Spec$^\#$~\cite{boogie} is a modern PVE for C$^\#$ that uses automated
theorem proving.  \escjava~\cite{escjava,cok2005eju} is an ESC for
JML-annotated~\cite{leavens1999jnd} Java code.  It produces VCs in the
Simplify~\cite{simplify} format and in the SMT format~\cite{smtlib}
for other automated theorem provers. It also generates VCs for the Coq
interactive theorem prover~\cite{coq}.

Whether an ESC is considered a PVE or not depends chiefly
on how well integrated it is with the editor. \escjava\
is integrated into Eclipse using a plugin. Spec$^\#$ is
more tightly integrated into Visual Studio using a plugin.
Work on incremental compilation~\cite{schwartz1984icm}
suggests that an even tighter integration leads to important
performance benefits.

There are two improvements that we will try in the near 
future. One is to prune the DSA graph. The other is
to modify Fx7~\cite{fx7} to produce a formula weaker
than the query but still \unsat, and use that to prune
subsequent queries. Another idea that is worth exploring
is to integrate pruning more tightly not with the ESC but 
instead with the proving process. For example, we could
save the relevance of specific axioms in the old proof,
so they can be prioritized while searching for a proof 
of the new query.

To assess the effectiveness of these improvements we need
a better benchmark. The amount of JML-annotated
Java is still modest. Moreover, code from the version 
control history is not appropriate because the commit 
cycle is typically much longer that the duration between 
two invocations of \escjava. Therefore we need to collect 
such data ourselves and this is a time consuming effort. 
Such a benchmark would hopefully nicely complement the 
existing (very useful) Boogie benchmarks and SMT-COMP 
benchmarks~\cite{smtlib}. A theoretical analysis seems
to require a good model for the type of queries that
are produced as verification conditions.

An idea very similar to the one explored in this paper did
lead to interesting results in model checking~\cite{henzinger2004emc},
the so called \emph{extreme model checking}. Model checking
is sometimes used together with unit testing and therefore
it is run often on code with minor modifications.
Therefore, it is natural to take advantage of the results
of previous runs.

\section{Conclusion}

We described the typical usage pattern of automated theorem
proving in extended static checking and two approaches that
exploit it to improve performance. We gave a detailed solution 
that processes first order formulas. The implementation
is a part of the Fx7 theorem prover~\cite{fx7}. It was tested
on queries generated by \escjava, without requiring any
modifications to the latter. The other approach, working on
the intermediate representation of the extended static checker,
promises to be more efficient but requires a tighter integration
of the prover with the checker.

The first part of the solution is a heuristic that, given
two formulas, finds which sub-terms of one formula correspond
to which sub-terms of the other. This heuristic may prove to be
a useful technique in solving related problems since it
performs well and there is ample room for tuning. The second
part of the solution is a formula pruning algorithm. This
algorithm is proven correct, and part of the proof is 
mechanically verified. Its efficiency is reasonable because
of the use of hash-consing and because formulas are normalized
with respect to commutative operators. The pruned formulas 
are clearly easier to prove.

\bigskip\penalty-500\noindent\textbf{Acknowledgements.}
This work is partly funded by the Information Society Technologies 
program of the European Commission, Future and Emerging
Technologies under the IST-2005-015905 MOBIUS project. 
The article contains only the authors' views and the Community
is not liable for any use that may be made of the information
therein. The second author is partially supported by Polish 
Ministry of Science and Education grant 3 T11C 042 30.

The authors would like to thank Joseph Kiniry, Mikol\'a\v s
Janota, and Fintan Fairmichael for their detailed feedback
on a draft of this article. The authors would also like to
thank the anonymous reviewers who pointed out that a formal
analysis of the performance gains is needed. We will try to
include such an analysis once the work progresses.

\bibliography{smt}


\end{document}